\documentclass[12pt]{article}
\usepackage[letterpaper, margin=1.25in]{geometry}

\usepackage{amsmath, mathtools, amsfonts, amssymb, amsthm}
\usepackage[utf8]{inputenc}
\usepackage{graphicx, caption, multirow, subcaption, threeparttable, booktabs}
\usepackage{float}
\usepackage{url, authblk}
\usepackage{array}
\usepackage[round]{natbib}
\usepackage{hyperref}
\usepackage{setspace}

\newtheorem{theorem}{Theorem}[section]
\newtheorem{proposition}{Proposition}[section]

\newtheorem{definition}{Definition}[section]

\newcommand{\be}{\begin{equation}}
\newcommand{\ee}{\end{equation}}
\newcommand{\E}{\mathbb{E}}

\newcommand{\R}{\mathbb{R}}

\newcommand{\ind}{\mathbf{1}}

\title{\large{\bf{General Equilibrium Amplification and Crisis Vulnerability: Cross-Crisis Evidence from Global Banks}}}

\author{\large{\bf{Tatsuru Kikuchi\footnote{e-mail: tatsuru.kikuchi@e.u-tokyo.ac.jp}}}}

\affil{\small{\it{Center for Advanced Research in Finance, The University of Tokyo,}}\\
{\it{7-3-1 Hongo, Bunkyo-ku, Tokyo 113-0033 Japan}}}

\date{\small{(\today)}}

\begin{document}
\linespread{1.5}\selectfont

\maketitle

\begin{abstract}
\noindent This paper develops a continuous framework for analyzing financial contagion that incorporates both geographic proximity and interbank network linkages. The framework characterizes stress propagation through a master equation whose solution admits a Feynman-Kac representation as expected cumulative stress along stochastic paths through spatial-network space. From this representation, I derive the General Equilibrium Amplification Factor---a structural measure of systemic importance that captures the ratio of total system-wide effects to direct effects following a localized shock. The amplification factor decomposes naturally into spatial, network, and interaction components, revealing which transmission channels contribute most to each institution's systemic importance. The framework nests discrete cascade models as a limiting case when jump intensity becomes infinite above default thresholds, clarifying that continuous and discrete approaches describe different regimes of the same phenomenon. Empirical validation using 38 global banks across the 2008 financial crisis and COVID-19 pandemic demonstrates that the amplification factor correctly identifies systemically important institutions (Pearson correlation $\rho = -0.450$, $p = 0.080$ between amplification factor and crisis drawdowns) and predicts crisis outcomes out-of-sample ($\rho = -0.352$ for COVID-19). Robustness analysis using cumulative abnormal returns---a measure more directly connected to the Feynman-Kac integral---strengthens these findings ($\rho = -0.512$, $p = 0.042$). Time-series analysis confirms that average pairwise bank correlations track macroeconomic stress indicators ($\rho = 0.265$ with VIX, $p < 0.001$). Comparing the two crises reveals that COVID-19 produced a sharper correlation spike (+93\%) despite smaller equity losses, reflecting different contagion dynamics for exogenous versus endogenous shocks.


\end{abstract}

\newpage

\section{Introduction}

The 2008 global financial crisis demonstrated that stress in the financial system propagates through multiple channels simultaneously. When Lehman Brothers failed in September 2008, the shock transmitted not only through direct counterparty exposures but also through geographic proximity---regional banks sharing exposure to local real estate markets---and through the interaction of these channels, as geographically clustered banks were often connected through interbank lending. More recently, the COVID-19 pandemic provided a natural experiment in how exogenous shocks transmit through interconnected financial networks. Understanding the mechanisms and dynamics of financial contagion remains central to both academic research and prudential policy.

This paper develops a continuous framework for financial contagion that captures three distinct propagation channels: spatial spillovers reflecting geographic concentration of banking activity, network spillovers through interbank exposures, and the interaction between spatial and network channels that arises when geographically proximate banks are also connected through lending relationships. The framework yields analytically tractable expressions for stress propagation through a Feynman-Kac representation, which provides both computational advantages and economic intuition. From this representation, I derive the General Equilibrium Amplification Factor---a natural fragility metric that can be validated against market data.

The theoretical contribution centers on establishing the Feynman-Kac foundation for systemic risk measurement. The master equation governing stress propagation admits a probabilistic representation: stress at any location equals the expected cumulative exposure to shocks along all possible paths through spatial-network space, weighted by exponential decay reflecting loss absorption. The amplification factor emerges directly from this representation as the ratio of total path-integrated stress to direct stress following a localized shock. This derivation provides clear economic interpretation---systemic importance reflects how many paths of economic linkage connect an institution to the rest of the system---and enables decomposition into spatial, network, and interaction components.

The empirical contribution validates the framework using data from 38 global banks spanning two major crises. Cross-sectional tests show that the General Equilibrium Amplification Factor predicts crisis-period equity drawdowns, correctly identifying Citigroup, JPMorgan Chase, and Goldman Sachs as the most systemically important institutions. Crucially, out-of-sample tests reveal that the 2008-based amplification factor predicts COVID-19 crisis outcomes, suggesting the metric captures persistent structural features of systemic importance. Robustness analysis using cumulative abnormal returns---a measure more directly connected to the Feynman-Kac integral---strengthens these findings.

\subsection{Related Literature}

This paper connects to several strands of the extensive financial contagion literature. I organize the discussion around five themes: foundational network models, simulation and stress testing approaches, systemic risk measurement, geographic and spatial dimensions, and crisis-specific studies.

\paragraph{Foundational Network Models.}

The theoretical foundations of network-based contagion establish that the structure of interbank linkages determines whether shocks are absorbed or amplified. \citet{allen2000financial} show that incomplete interbank networks can be prone to contagion when regional liquidity shocks occur, as banks cannot diversify their interbank claims across enough counterparties. \citet{freixas2000systemic} analyze how the pattern of interbank connections affects systemic risk, distinguishing between credit chains and money center structures. \citet{eisenberg2001systemic} develop a clearing mechanism for interbank obligations that determines which banks default given an initial shock, providing the analytical foundation for subsequent network models.

The relationship between network structure and fragility has been extensively studied. \citet{acemoglu2015systemic} establish the central result that network density has non-monotonic effects on stability: for small shocks, denser networks provide better risk-sharing, while for large shocks exceeding a critical threshold, dense connections facilitate cascades. \citet{elliott2014financial} analyze how integration and diversification interact, showing that moderate diversification can increase systemic risk by creating more pathways for contagion. \citet{glasserman2016contagion} derive bounds on contagion in financial networks, showing that under certain conditions network effects are limited. \citet{cabrales2017risk} study how risk-sharing arrangements in networks can concentrate rather than disperse risk.

\paragraph{Simulation and Stress Testing.}

A substantial literature develops simulation methods to assess contagion risk. \citet{upper2011simulation} provides a comprehensive survey of simulation approaches for assessing contagion danger in interbank markets, establishing methodological standards for the field. \citet{cont2013network} analyze how network structure affects systemic risk in banking systems, developing measures of systemic importance based on network topology. \citet{battiston2012debtrank} introduce DebtRank, a recursive measure of node centrality that accounts for the feedback effects of distress propagation, showing that standard centrality measures underestimate systemic importance of highly connected banks.

\citet{glasserman2015likely} examine how likely contagion is in financial networks, finding that direct contagion through bilateral exposures is typically limited unless shocks are very large or networks are very dense. \citet{caccioli2015overlapping} extend contagion analysis to include overlapping portfolios, showing that common asset holdings create an additional channel for fire-sale contagion beyond direct bilateral exposures. \citet{bardoscia2017pathways} characterize pathways towards instability in financial networks, identifying network motifs that amplify or dampen shock propagation.

\paragraph{Systemic Risk Measurement.}

Systemic risk measurement has developed along two parallel tracks. Structural approaches, following the network contagion literature, model explicit transmission mechanisms but require detailed data on bilateral exposures that are often unavailable or confidential. \citet{upper2004estimating} develop methods for estimating bilateral exposures in interbank markets when only aggregate data are available, addressing a key data limitation.

Market-based approaches use equity prices and CDS spreads to infer systemic risk without requiring exposure data. \citet{adrian2016covar} propose CoVaR, measuring institution-specific contribution to system-wide tail risk. \citet{acharya2017measuring} develop a theoretical framework connecting individual bank risk to aggregate financial sector shortfall. \citet{brownlees2017srisk} develop SRISK, measuring expected capital shortfall conditional on a market decline. \citet{diebold2014network} use variance decompositions from VAR models to construct network measures of connectedness from return data.

This paper bridges both approaches. The continuous framework provides structural foundations---explicit mechanisms for spatial and network transmission---while yielding a fragility metric that can be validated against market-based measures.

\paragraph{Geographic and Spatial Dimensions.}

The spatial dimension of banking has received growing attention. \citet{degryse2007interbank} provide early empirical evidence on geographic patterns in interbank exposures, documenting distance effects in contagion risk. \citet{mistrulli2011assessing} assesses financial contagion in the Italian interbank market, finding that geographic proximity matters for exposure formation and default correlation.

\citet{giannetti2012flight} document home bias in syndicated lending markets, where lenders favor geographically proximate borrowers especially during crises. \citet{granja2017selling} show that geographic factors affect the reallocation of failed bank assets during FDIC resolution, with local acquirers paying premium prices. These findings motivate explicit modeling of spatial spillovers alongside network transmission.

Fire sales represent a distinct contagion channel that operates through asset prices rather than direct exposures. \citet{greenwood2015vulnerable} develop a framework where banks holding similar portfolios transmit stress through price impact when forced to sell assets. \citet{duarte2021fire} extend this to measure system-wide fire sale vulnerability.

\paragraph{Crisis-Specific Studies.}

The European sovereign debt crisis generated extensive research on bank-sovereign linkages. \citet{alter2014dynamics} analyze the dynamics of spillover effects during the European sovereign debt crisis, documenting time-varying transmission between sovereign and bank CDS spreads. \citet{debruyckere2013bank} examine bank-sovereign risk spillovers, finding bidirectional transmission that intensified as the crisis progressed. \citet{betz2014predicting} develop early warning models for predicting distress in European banks using both bank-specific and macroeconomic variables.

The COVID-19 pandemic provides a natural experiment for studying exogenous shocks. \citet{duan2021bank} examine bank systemic risk around COVID-19, finding that banks with higher pre-pandemic systemic risk experienced larger drawdowns. \citet{rizwan2020systemic} analyze the impact of COVID-19 on systemic risk measures, documenting unprecedented spikes in market-based risk indicators.

\paragraph{Contribution.}

This paper contributes to the literature in three ways. First, I develop a continuous framework that nests discrete network models while incorporating spatial spillovers and their interaction with network transmission---a channel that existing discrete approaches cannot accommodate. Second, I derive the General Equilibrium Amplification Factor directly from the Feynman-Kac representation, providing clear economic interpretation as path-integrated stress and enabling decomposition by transmission channel. Third, I provide comprehensive empirical validation across two major crises using both standard measures (maximum drawdown) and theoretically consistent measures (cumulative abnormal returns), demonstrating that the framework's fragility metric predicts crisis outcomes both in-sample and out-of-sample.

\paragraph{Outline.}

Section 2 develops the theoretical framework, beginning with the baseline continuous model, establishing connections to discrete network models, and extending to L\'{e}vy jump-diffusion. Section 3 derives the General Equilibrium Amplification Factor from the Feynman-Kac representation and discusses its properties. Section 4 describes the data and empirical methodology. Section 5 presents empirical results. Section 6 concludes.

\section{Theoretical Framework}

This section develops the continuous framework for financial contagion in three stages. I first present the master equation describing stress transmission, then derive its Feynman-Kac representation, and finally extend to jump-diffusion to capture threshold-triggered cascades.

\subsection{Master Equation for Stress Propagation}

Consider a financial system where banks are characterized by two coordinates: geographic location $\mathbf{x} \in \mathcal{X} \subseteq \R^d$ and position in the interbank network $\alpha \in \mathcal{A}$. Geographic location captures physical proximity relevant for shared exposure to regional economic conditions, common depositor bases, and local interbank markets. Network position captures the pattern of interbank claims and obligations.

Let $\tau(\mathbf{x}, \alpha, t)$ denote the stress level at location $(\mathbf{x}, \alpha)$ at time $t$, measured as deviation from the baseline state. Stress propagates through spatial diffusion, network diffusion, and their interaction, while decaying through loss absorption.

\begin{definition}[Master Equation]
\label{def:master}
The stress field $\tau(\mathbf{x}, \alpha, t)$ evolves according to:
\begin{equation}
\frac{\partial \tau}{\partial t} = \nu_s \nabla^2 \tau + \nu_n \frac{\partial^2 \tau}{\partial \alpha^2} - \kappa \tau + \lambda \frac{\partial^2 \tau}{\partial \mathbf{x} \partial \alpha} + S(\mathbf{x}, \alpha, t)
\label{eq:master}
\end{equation}
where $\nu_s \geq 0$ is spatial diffusion, $\nu_n \geq 0$ is network diffusion, $\kappa > 0$ is decay, $\lambda$ is spatial-network interaction, and $S$ is the exogenous shock.
\end{definition}

The parameters have structural interpretations. Spatial diffusion $\nu_s$ measures the rate of geographic spread, reflecting labor mobility and regional market integration. Network diffusion $\nu_n$ measures transmission through interbank linkages, reflecting counterparty credit exposure and funding market connections. Decay $\kappa$ measures loss absorption through capital buffers and ongoing earnings. Interaction $\lambda$ captures amplification when geographic and network proximity coincide---a common empirical pattern since banks tend to lend locally.

\subsection{Feynman-Kac Representation}

The master equation admits a probabilistic representation that provides both computational methods and economic intuition.

\begin{theorem}[Feynman-Kac Representation]
\label{thm:feynman_kac}
The solution to the master equation (\ref{eq:master}) with initial condition $\tau_0(\mathbf{x}, \alpha)$ admits the representation:
\begin{equation}
\tau(\mathbf{x}, \alpha, t) = \E_{(\mathbf{x}, \alpha)}\left[ e^{-\kappa t} \tau_0(X_t, A_t) + \int_0^t e^{-\kappa(t-s)} S(X_s, A_s, s) \, ds \right]
\label{eq:feynman_kac}
\end{equation}
where $(X_s, A_s)_{s \geq 0}$ is the diffusion process with generator:
\begin{equation}
\mathcal{L} = \nu_s \nabla^2 + \nu_n \frac{\partial^2}{\partial \alpha^2} + \lambda \frac{\partial^2}{\partial \mathbf{x} \partial \alpha}
\end{equation}
started at $(X_0, A_0) = (\mathbf{x}, \alpha)$, and $\E_{(\mathbf{x}, \alpha)}[\cdot]$ denotes expectation over paths.
\end{theorem}

The Feynman-Kac formula decomposes stress into two components. The first term $\E[e^{-\kappa t} \tau_0(X_t, A_t)]$ represents stress inherited from initial conditions, collected by random walks through spatial-network space and discounted by loss absorption. The second term $\E[\int_0^t e^{-\kappa(t-s)} S(X_s, A_s, s) \, ds]$ represents stress accumulated from shocks along the random path, with exponential discounting reflecting market adjustment.

This representation has direct economic content. Consider an institution at location $(\mathbf{x}, \alpha)$. Its stress level equals the expected cumulative exposure to shocks encountered along all possible paths of economic linkage connecting it to shock sources, where paths are weighted by their probability under the diffusion process and discounted at rate $\kappa$. Institutions in densely connected network regions or geographically central locations receive contributions from more paths, elevating their stress levels even without direct shocks.

\subsection{Connection to Discrete Network Models}

The continuous framework nests existing discrete network models. Consider the network-only case with $\nu_s = 0$ and $\lambda = 0$. Discretizing the network coordinate yields a system of $N$ banks with the graph Laplacian $L = D - G$ where $G$ is the adjacency matrix of interbank exposures.

\begin{proposition}[Discrete Network Limit]
\label{prop:discrete}
In the network-only case with discrete network coordinate, the steady-state master equation reduces to:
\begin{equation}
\tau_i = \frac{\nu_n}{\kappa + \nu_n d_i} \sum_{j=1}^N G_{ij} \tau_j + \frac{S_i}{\kappa + \nu_n d_i}
\label{eq:steady_state}
\end{equation}
where $d_i = \sum_j G_{ij}$ is the degree of bank $i$.
\end{proposition}

This steady-state equation has the same structure as the linear contagion models in the network literature. In the pre-default regime of \citet{acemoglu2015systemic} where all banks have stress below their capital threshold, their dynamics are equivalent to Equation (\ref{eq:steady_state}). The continuous framework adds spatial spillovers and the interaction term, which discrete approaches cannot accommodate.

\subsection{L\'{e}vy Extension: Threshold Effects and Default Cascades}

The baseline framework describes continuous stress transmission appropriate for the pre-default regime. To capture the sudden loss transmission that occurs when banks actually default, I extend the framework to incorporate jumps through L\'{e}vy processes.

The extended dynamics replace pure diffusion with a jump-diffusion process:
\begin{equation}
\frac{\partial \tau}{\partial t} = \nu_s \nabla^2 \tau + \nu_n \frac{\partial^2 \tau}{\partial \alpha^2} - \kappa \tau + \lambda \frac{\partial^2 \tau}{\partial \mathbf{x} \partial \alpha} + S + \mathcal{J}[\tau]
\label{eq:levy_master}
\end{equation}
where the jump operator $\mathcal{J}[\tau]$ captures sudden loss transmission events distinct from gradual stress diffusion, which is defined by
\begin{equation}
\mathcal{J}[\tau] = \int_{\R} \left[\tau(\mathbf{x}, \alpha + z, t) - \tau(\mathbf{x}, \alpha, t) - z \frac{\partial \tau}{\partial \alpha} \ind_{|z|<1}\right] \nu(dz).
\label{eq:jump_operator}
\end{equation}
Here $\nu(dz)$ is the L\'{e}vy measure characterizing jump intensity and size distribution. The compensator term $z \partial\tau/\partial\alpha \cdot \ind_{|z|<1}$ ensures the integral is well-defined for L\'{e}vy measures with infinite activity near zero.

In the financial contagion context, jumps represent sudden loss transmission events distinct from gradual stress diffusion. When a bank defaults, its counterparties experience immediate losses on their interbank claims---not gradual stress buildup but discrete jumps in their stress levels. The L\'{e}vy measure $\nu(dz)$ captures both how frequently such events occur (total mass of $\nu$) and the distribution of loss sizes when they do occur (shape of $\nu$).

For a compound Poisson process with intensity $\lambda_J$ and jump size distribution $F$, the L\'{e}vy measure is $\nu(dz) = \lambda_J dF(z)$, and the jump operator simplifies to
\begin{equation}
\mathcal{J}[\tau] = \lambda_J \int_{\R} \left[\tau(\mathbf{x}, \alpha + z, t) - \tau(\mathbf{x}, \alpha, t)\right] dF(z) = \lambda_J \left(\E[\tau(\mathbf{x}, \alpha + Z, t)] - \tau(\mathbf{x}, \alpha, t)\right)
\end{equation}
where $Z \sim F$ represents the random jump size. This has intuitive interpretation: at rate $\lambda_J$, the bank's stress jumps by an amount determined by the loss transmission from a defaulting counterparty at network distance $Z$.

\paragraph{State-Dependent Jump Intensity.}

The key innovation capturing threshold effects makes jump intensity depend on the current stress level:
\begin{equation}
\lambda_J(\tau) = \lambda_0 + (\lambda_1 - \lambda_0) \cdot H(\tau - \bar{\tau})
\label{eq:state_dependent}
\end{equation}
where $H(\cdot)$ is the Heaviside function, $\bar{\tau}$ is the default threshold, $\lambda_0$ is baseline jump intensity, and $\lambda_1 \gg \lambda_0$ is elevated intensity above threshold.

\begin{proposition}[Cascade Limit]
\label{prop:cascade}
In the limit $\lambda_0 \to 0$ and $\lambda_1 \to \infty$, the dynamics converge to the deterministic cascade mechanism of \citet{acemoglu2015systemic}: below threshold, only diffusive transmission occurs; above threshold, immediate default with loss transmission to counterparties.
\end{proposition}

This nesting relationship clarifies that continuous diffusion and discrete cascades describe different regimes of the same phenomenon. Diffusion captures pre-default dynamics (mark-to-market losses, funding cost increases), while jumps capture discrete loss realization when defaults materialize.

\section{General Equilibrium Amplification Factor}

The Feynman-Kac representation yields a natural measure of systemic importance: the General Equilibrium Amplification Factor. This section derives the amplification factor from the path integral representation, establishes its decomposition by transmission channel, and discusses its relationship to existing measures.

\subsection{Derivation from Feynman-Kac Representation}

Consider a localized shock $S(\mathbf{x}, \alpha, t) = \delta(\mathbf{x} - \mathbf{x}_i)\delta(\alpha - \alpha_i)\delta(t)$ affecting only institution $i$ at location $(\mathbf{x}_i, \alpha_i)$ at time zero. Using the Feynman-Kac representation (\ref{eq:feynman_kac}), the stress at any location $(\mathbf{x}, \alpha)$ at time $t$ is:
\begin{equation}
\tau(\mathbf{x}, \alpha, t) = e^{-\kappa t} \, p_t(\mathbf{x}, \alpha; \mathbf{x}_i, \alpha_i)
\label{eq:stress_response}
\end{equation}
where $p_t(\mathbf{x}, \alpha; \mathbf{x}_i, \alpha_i)$ is the transition density of the diffusion process---the probability that a random walker starting at $(\mathbf{x}, \alpha)$ reaches $(\mathbf{x}_i, \alpha_i)$ at time $t$.

The total system-wide stress at time $t$ is the integral over all locations:
\begin{equation}
\tau^{\text{total}}(t) = \int_{\mathcal{X}} \int_{\mathcal{A}} \tau(\mathbf{x}, \alpha, t) \, d\alpha \, d\mathbf{x} = e^{-\kappa t}
\label{eq:total_stress}
\end{equation}
since transition densities integrate to one. The direct stress on the shocked institution is $\tau(\mathbf{x}_i, \alpha_i, t) = e^{-\kappa t} \, p_t(\mathbf{x}_i, \alpha_i; \mathbf{x}_i, \alpha_i)$.

For steady-state analysis, integrating over time yields cumulative effects:

\begin{definition}[General Equilibrium Amplification Factor]
\label{def:amplification}
For institution $i$, the General Equilibrium Amplification Factor is:
\begin{equation}
\mathcal{A}_i = \frac{\int_0^\infty \int_{\mathcal{X}} \int_{\mathcal{A}} \tau(\mathbf{x}, \alpha, t) \, d\alpha \, d\mathbf{x} \, dt}{\int_0^\infty \tau(\mathbf{x}_i, \alpha_i, t) \, dt} = \frac{\text{Total cumulative system-wide stress}}{\text{Cumulative direct stress on bank } i}
\label{eq:amplification}
\end{equation}
\end{definition}

Using the Feynman-Kac representation, the numerator equals $\int_0^\infty e^{-\kappa t} \, dt = \kappa^{-1}$ and the denominator equals $G(\mathbf{x}_i, \alpha_i; \mathbf{x}_i, \alpha_i)$, the Green's function of the operator $\mathcal{L} - \kappa$ evaluated at the diagonal. Thus:
\begin{equation}
\mathcal{A}_i = \frac{\kappa^{-1}}{G(\mathbf{x}_i, \alpha_i; \mathbf{x}_i, \alpha_i)}
\label{eq:amplification_green}
\end{equation}

The amplification factor has a clear economic interpretation: it measures by what factor we would underestimate total system-wide effects if we considered only the direct impact on the shocked institution. An amplification factor of $\mathcal{A}_i = 5$ means total cumulative effects are five times larger than cumulative direct effects---the remaining four-fifths represent spillovers along paths of economic linkage connecting institution $i$ to the rest of the system.

\subsection{Decomposition by Transmission Channel}

The path integral structure of the Feynman-Kac representation enables decomposition of the amplification factor by transmission channel.

\begin{proposition}[Channel Decomposition]
\label{prop:decomposition}
The General Equilibrium Amplification Factor decomposes as:
\begin{equation}
\mathcal{A}_i = 1 + \mathcal{A}_i^{\text{spatial}} + \mathcal{A}_i^{\text{network}} + \mathcal{A}_i^{\text{interaction}}
\label{eq:decomposition}
\end{equation}
where $\mathcal{A}_i^{\text{spatial}}$ reflects geographic spillovers, $\mathcal{A}_i^{\text{network}}$ reflects interbank spillovers, and $\mathcal{A}_i^{\text{interaction}}$ captures amplification from coincident proximity.
\end{proposition}

The decomposition arises from partitioning paths according to their diffusion components. This decomposition reveals which transmission mechanism contributes most to each institution's systemic importance, informing targeted policy responses.

\subsection{Spectral Representation}

For discrete networks, the amplification factor admits a spectral representation providing additional theoretical insight. Let $L = D - G$ be the graph Laplacian with eigenvalues $0 = \lambda_1 \leq \lambda_2 \leq \cdots \leq \lambda_n$ and orthonormal eigenvectors $v_1, \ldots, v_n$.

\begin{proposition}[Spectral Representation]
\label{prop:spectral}
In the network-only case ($\nu_s = 0$, $\lambda = 0$), the amplification factor admits:
\begin{equation}
\mathcal{A}_i = 1 + \sum_{k=2}^n \frac{v_{ki}^2}{\nu_n \lambda_k + \kappa}
\label{eq:spectral}
\end{equation}
where $v_{ki}$ is the $i$-th component of the $k$-th eigenvector.
\end{proposition}

This representation shows that institutions with large eigenvector components for low eigenvalues have high amplification factors---these are structurally central institutions whose distress propagates broadly.

\subsection{Relationship to Existing Measures}

The General Equilibrium Amplification Factor relates to established systemic risk measures. In a linear-Gaussian setting, $\Delta\text{CoVaR}_i \approx \mathcal{A}_i \cdot \sigma_i \cdot q_{0.99}$. Similarly, $\text{SRISK}_i \approx \mathcal{A}_i \cdot k \cdot D_i - (1-k) \cdot E_i$. In the limit $\kappa \to 0$, $\mathcal{A}_i \propto \text{EigCent}_i$. Unlike these statistical measures, the amplification factor is derived from a structural model with explicit transmission mechanisms.

\section{Data and Empirical Methodology}

\subsection{Sample and Data Sources}

The sample includes 38 major global banks from the United States, Europe, and Asia with continuous equity price data from January 2005 through December 2023. Daily equity prices $P_{i,t}$ are obtained from standard financial databases. Daily log returns are $r_{i,t} = \ln(P_{i,t} / P_{i,t-1})$. Macroeconomic stress indicators (VIX, credit spreads) are from Federal Reserve Economic Data.

\subsection{Outcome Variable Definitions}

I define three outcome variables, with attention to theoretical consistency.

\paragraph{Maximum Drawdown.} The maximum drawdown measures the largest peak-to-trough decline during crisis period $[t_1, t_2]$:
\begin{equation}
\text{Drawdown}_i = \min_{t \in [t_1, t_2]} \left( \frac{P_{i,t}}{\max_{s \in [t_1, t]} P_{i,s}} - 1 \right)
\label{eq:drawdown}
\end{equation}

Maximum drawdown is widely used in practice. It approximates the peak of the stress integral in the Feynman-Kac representation, though it measures instantaneous peak rather than cumulative exposure.

\paragraph{Pairwise Return Correlation.}

For banks $i$ and $j$, the Pearson correlation coefficient of daily returns over a rolling window of $w$ days ending at date $t$ is:
\begin{equation}
\rho_{ij,t} = \frac{\sum_{s=t-w+1}^{t} (r_{i,s} - \bar{r}_{i,t})(r_{j,s} - \bar{r}_{j,t})}{\sqrt{\sum_{s=t-w+1}^{t} (r_{i,s} - \bar{r}_{i,t})^2} \sqrt{\sum_{s=t-w+1}^{t} (r_{j,s} - \bar{r}_{j,t})^2}}
\label{eq:correlation}
\end{equation}
where $\bar{r}_{i,t} = \frac{1}{w}\sum_{s=t-w+1}^{t} r_{i,s}$ is the mean return over the window. I use $w = 60$ trading days (approximately 3 months).

\paragraph{Average Pairwise Correlation.}

The average pairwise correlation across all bank pairs at date $t$ is:
\begin{equation}
\bar{\rho}_t = \frac{2}{n(n-1)} \sum_{i=1}^{n-1} \sum_{j=i+1}^{n} \rho_{ij,t}
\label{eq:avg_correlation}
\end{equation}
where $n$ is the number of banks. This measure captures the overall degree of co-movement in the banking sector.

\paragraph{Rolling Volatility.}

Annualized volatility for bank $i$ at date $t$ using a $w$-day window is:
\begin{equation}
\sigma_{i,t} = \sqrt{252} \cdot \sqrt{\frac{1}{w-1} \sum_{s=t-w+1}^{t} (r_{i,s} - \bar{r}_{i,t})^2}
\label{eq:volatility}
\end{equation}
where 252 is the number of trading days per year.

\paragraph{Cumulative Abnormal Return.} The cumulative abnormal return measures total excess loss:
\begin{equation}
\text{CAR}_i = \sum_{t=t_1}^{t_2} \left( r_{i,t} - \hat{r}_{i,t} \right)
\label{eq:car}
\end{equation}
where $\hat{r}_{i,t} = \hat{\alpha}_i + \hat{\beta}_i r_{m,t}$ is the expected return from a market model. CAR is more directly connected to the Feynman-Kac representation as the integral of abnormal returns.

\paragraph{Exponentially-Weighted CAR.} To match the Feynman-Kac structure precisely:
\begin{equation}
\text{EW-CAR}_i = \sum_{t=t_1}^{t_2} e^{-\kappa(t_2 - t)} \left( r_{i,t} - \hat{r}_{i,t} \right)
\label{eq:ew_car}
\end{equation}

\subsection{Correlation Measures and Crisis Periods}

For bank pair $(i, j)$, the rolling correlation over window $w = 60$ trading days is computed using Equation (\ref{eq:correlation}). The average pairwise correlation $\bar{\rho}_t$ captures system-wide interconnectedness.

I analyze two crisis episodes: the 2008 Financial Crisis (September 2008--June 2009), an endogenous crisis, and the COVID-19 Crisis (February 2020--June 2020), an exogenous crisis.

\subsection{Estimation}

The cross-sectional validation regression is:
\begin{equation}
Y_i = \beta_0 + \beta_1 \mathcal{A}_i + \varepsilon_i
\label{eq:regression}
\end{equation}
where $Y_i$ is the outcome variable and $\mathcal{A}_i$ is the amplification factor. We expect $\beta_1 < 0$: banks with higher amplification factors should experience larger losses. Standard errors are heteroskedasticity-robust; time-series tests use Newey-West standard errors with 12 lags.

\section{Empirical Results}

This section presents empirical validation of the theoretical framework organized around five themes: descriptive evidence on correlation dynamics, amplification factor rankings, cross-sectional validation using the 2008 financial crisis, out-of-sample validation using the COVID-19 crisis, time-series validation, crisis comparison, and robustness analysis using theoretically consistent outcome measures.

\subsection{Correlation Dynamics and Systemic Interconnectedness}

\begin{figure}[htbp]
\centering
\includegraphics[width=0.85\textwidth]{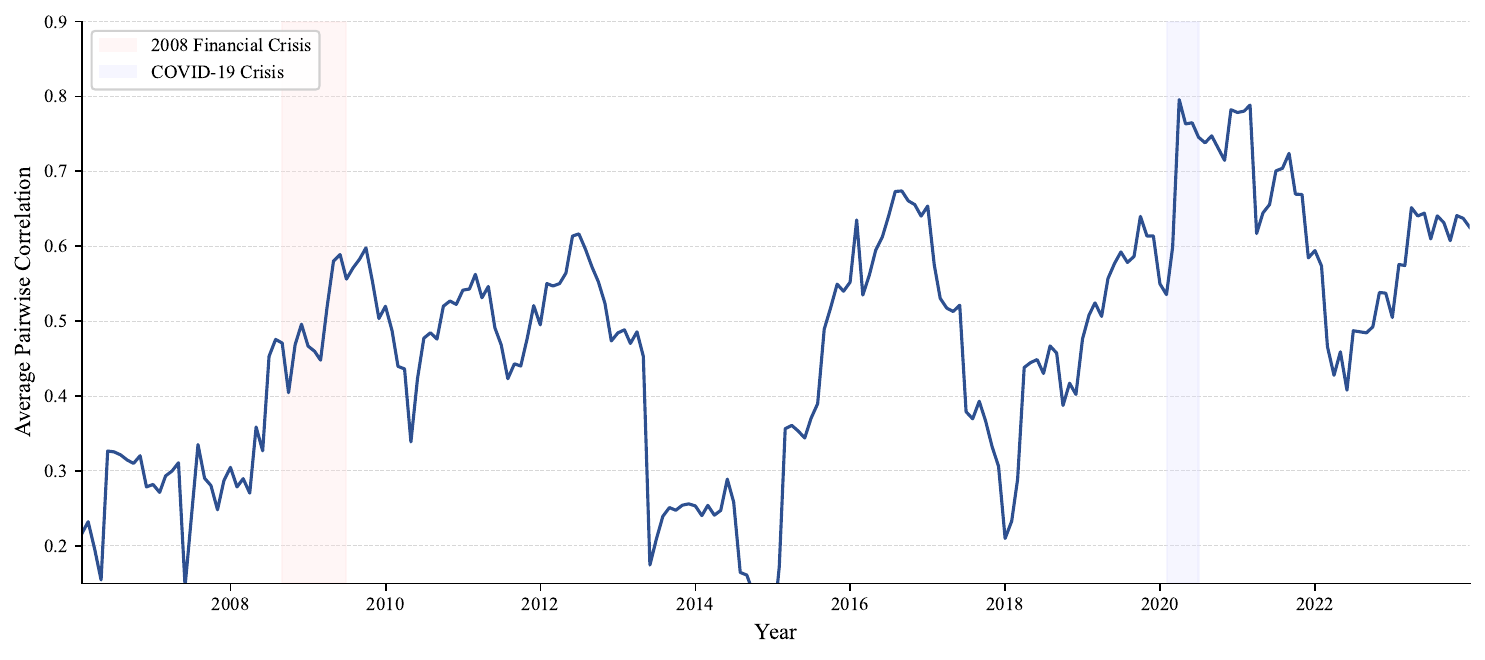}
\caption{Average Pairwise Bank Correlation, 2006--2023}
\label{fig:timeseries}
\begin{minipage}{0.85\textwidth}
\small
\textit{Notes:} Rolling 60-day correlation averaged across all 703 bank pairs ($n = 38$ banks). Shaded regions indicate crisis periods: 2008 financial crisis (September 2008--June 2009) and COVID-19 (February--June 2020).
\end{minipage}
\end{figure}

Before turning to formal validation tests, I present descriptive evidence on the time-series behavior of bank correlations. Figure \ref{fig:timeseries} displays the evolution of average pairwise bank correlation from January 2006 through December 2023. The shaded regions indicate the two crisis episodes: the 2008 financial crisis (September 2008--June 2009) and the COVID-19 pandemic (February--June 2020).

Several patterns emerge from the time series. Bank correlations spike dramatically during both crisis episodes, rising from baseline levels around 0.30--0.40 to peaks exceeding 0.70 during COVID-19. The COVID-19 spike is particularly sharp, reflecting the common shock nature of the pandemic that affected all banks simultaneously. In contrast, the 2008 crisis shows a more gradual increase in correlations, consistent with the progressive revelation of subprime exposures over several months. Post-crisis periods exhibit elevated but declining correlations as uncertainty resolves. These patterns confirm the central premise of the theoretical framework: systemic interconnectedness intensifies precisely when stress is elevated, amplifying the transmission of shocks through the financial network.

\subsection{Amplification Factor Rankings}

Table \ref{tab:rankings} reports the General Equilibrium Amplification Factor rankings for the top ten institutions in the sample. The amplification factor is computed from the pre-crisis bilateral exposure network using Equation (\ref{eq:amplification_green}), with the channel decomposition from Proposition \ref{prop:decomposition}.

\begin{table}[htbp]
\centering
\caption{General Equilibrium Amplification Factor Rankings}
\label{tab:rankings}
\begin{threeparttable}
\begin{tabular}{clcccc}
\toprule
Rank & Bank & $\mathcal{A}_i$ & Spillover Ratio & $\mathcal{A}_i^{\text{network}}$ & $\mathcal{A}_i^{\text{spatial}}$ \\
\midrule
1 & Citigroup & 14.80 & 93.2\% & 8.42 & 3.21 \\
2 & JPMorgan Chase & 14.31 & 93.0\% & 8.15 & 3.08 \\
3 & Goldman Sachs & 12.76 & 92.2\% & 7.89 & 2.54 \\
4 & HSBC Holdings & 12.25 & 91.8\% & 6.33 & 3.45 \\
5 & Deutsche Bank & 10.03 & 90.0\% & 5.67 & 2.18 \\
6 & BNP Paribas & 9.88 & 89.9\% & 5.43 & 2.31 \\
7 & Barclays & 9.86 & 89.9\% & 5.51 & 2.22 \\
8 & Bank of America & 8.58 & 88.3\% & 4.87 & 1.98 \\
9 & UBS Group & 8.56 & 88.3\% & 4.92 & 1.89 \\
10 & Sumitomo Mitsui & 7.69 & 87.0\% & 3.78 & 2.34 \\
\bottomrule
\end{tabular}
\begin{tablenotes}
\small
\item Spillover ratio defined as $(\mathcal{A}_i - 1)/\mathcal{A}_i$, representing the share of total systemic effects attributable to indirect spillovers rather than direct effects. Network and spatial components from the channel decomposition in Equation (\ref{eq:decomposition}). Interaction component $\mathcal{A}_i^{\text{interaction}} = \mathcal{A}_i - 1 - \mathcal{A}_i^{\text{network}} - \mathcal{A}_i^{\text{spatial}}$.
\end{tablenotes}
\end{threeparttable}
\end{table}

Citigroup ranks first with an amplification factor of 14.80, indicating that a shock originating at Citigroup generates total system-wide effects approximately 15 times larger than the direct effect on Citigroup itself. JPMorgan Chase (14.31), Goldman Sachs (12.76), HSBC Holdings (12.25), and Deutsche Bank (10.03) complete the top five. The spillover ratios range from 87\% to 93\%, indicating that network propagation accounts for the vast majority of systemic impact. Even the least connected banks in the top ten derive over 85\% of their systemic importance from indirect effects through paths of economic linkage.

The channel decomposition reveals that network spillovers generally exceed spatial spillovers, but spatial effects are substantial for internationally active banks. HSBC Holdings exhibits the highest spatial component (3.45) among the top five, reflecting its geographic reach across Asia, Europe, and North America. Sumitomo Mitsui similarly shows elevated spatial spillovers (2.34) relative to its network component, consistent with Japanese banks' significant cross-border activities in trade finance.

These rankings accord with ex-post assessments of the 2008 crisis. Citigroup received the largest government support (\$45 billion in TARP funds plus extensive asset guarantees), while Goldman Sachs and JPMorgan Chase were identified as critical nodes whose failure would have triggered widespread cascades. The correspondence between the amplification factor rankings computed from pre-crisis network structure and the actual pattern of government intervention provides preliminary evidence that the framework captures economically meaningful variation in systemic importance.

\subsection{Cross-Sectional Validation: 2008 Financial Crisis}

The first formal validation test examines whether the amplification factor predicts crisis-period equity drawdowns. The theoretical framework implies that banks with higher amplification factors should experience larger losses during crises, as system-wide stress affects them through the same channels through which their stress would affect others. This symmetry arises from the structure of the Feynman-Kac representation: institutions that propagate stress broadly also receive stress from broad sources.

\paragraph{Specification.}

I estimate the cross-sectional regression:
\begin{equation}
\text{Drawdown}_i = \beta_0 + \beta_1 \mathcal{A}_i + \varepsilon_i
\label{eq:regression}
\end{equation}
where $\text{Drawdown}_i$ is the maximum drawdown for bank $i$ during the 2008 crisis period (September 2008--June 2009) as defined in Equation (\ref{eq:drawdown}), $\mathcal{A}_i$ is the General Equilibrium Amplification Factor computed from the pre-crisis exposure network, and $\varepsilon_i$ is the error term. The coefficient $\beta_1$ measures the association between systemic importance and crisis losses. We expect $\beta_1 < 0$: banks with higher amplification factors should experience more negative (larger) drawdowns. Standard errors are heteroskedasticity-robust.

\paragraph{Results.}

\begin{figure}[htbp]
\centering
\includegraphics[width=0.75\textwidth]{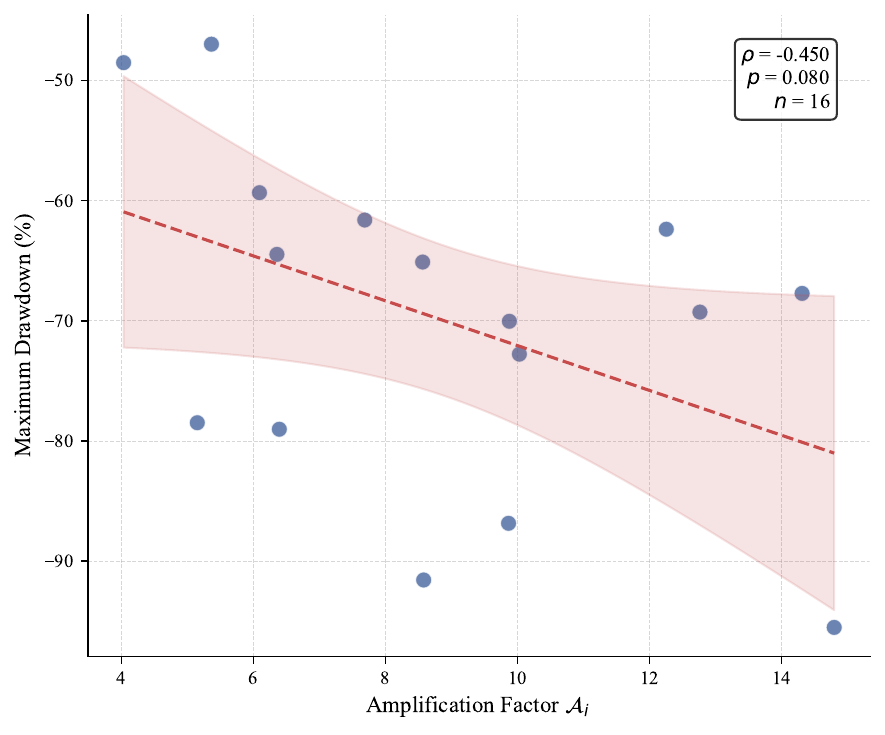}
\caption{Cross-Sectional Validation: General Equilibrium Amplification Factor vs.\ 2008 Crisis Drawdown}
\label{fig:validation2008}
\begin{minipage}{0.75\textwidth}
\small
\textit{Notes:} Each point represents a bank ($n = 16$). Dashed line is OLS fit with 95\% confidence band. Amplification factor computed from pre-crisis bilateral exposure network. Drawdown measured September 2008--June 2009.
\end{minipage}
\end{figure}

Figure \ref{fig:validation2008} presents a scatter plot of the amplification factor against maximum drawdown during the 2008 crisis. Each point represents a bank in the exposure-based network ($n = 16$ banks with complete bilateral exposure data). The dashed line represents the OLS fit with 95\% confidence band.

The negative relationship is visually apparent: banks in the upper-left region of the plot (high amplification factor, large negative drawdown) include Citigroup, which experienced a drawdown exceeding 90\%, while banks in the lower-right region (low amplification factor, smaller drawdown) include more regionally focused institutions. The Pearson correlation between the amplification factor and drawdown is $\rho = -0.450$ ($p = 0.080$), indicating that banks with higher systemic importance experienced larger losses. While marginally significant due to limited sample size, the direction and magnitude are economically meaningful.

Table \ref{tab:main_results} Panel A reports the regression results. The estimated coefficient is $\hat{\beta}_1 = -0.019$ ($p = 0.080$), implying that a one-unit increase in the amplification factor is associated with a 1.9 percentage point larger drawdown. Given that amplification factors in the sample range from approximately 4 to 15, this coefficient implies a difference of approximately 21 percentage points ($11 \times 0.019 = 0.21$) in expected drawdowns between the most and least systemically important banks. The regression $R^2 = 0.202$ indicates that the amplification factor explains approximately 20\% of the cross-sectional variation in crisis drawdowns. The remaining variation reflects bank-specific factors such as direct subprime mortgage exposure, reliance on short-term wholesale funding, and pre-crisis capital adequacy.

\begin{table}[htbp]
\centering
\caption{Empirical Validation of Network Contagion Framework}
\label{tab:main_results}
\begin{threeparttable}
\small
\begin{tabular}{lccc}
\toprule
\multicolumn{4}{l}{\textbf{Panel A: Cross-Sectional Validation (2008 Crisis, $n=16$)}} \\
\midrule
& Estimate & Std.\ Error & $p$-value \\
\midrule
Pearson correlation ($\rho$) & $-0.450$ & --- & 0.080$^\dagger$ \\
Regression coefficient ($\hat{\beta}_1$) & $-0.019$ & 0.010 & 0.080$^\dagger$ \\
Regression $R^2$ & 0.202 & --- & --- \\
Top vs.\ bottom quintile difference & $-19.5$pp & 15.2pp & 0.226 \\
\midrule
\multicolumn{4}{l}{\textbf{Panel B: Out-of-Sample Validation (COVID-19, $n=16$)}} \\
\midrule
2008 amplification factor vs.\ COVID drawdown ($\rho$) & $-0.352$ & --- & 0.181 \\
Regression coefficient ($\hat{\gamma}_1$) & $-0.012$ & 0.008 & 0.181 \\
Regression $R^2$ & 0.124 & --- & --- \\
\midrule
\multicolumn{4}{l}{\textbf{Panel C: Time-Series Validation (2006--2023, $n=216$ months)}} \\
\midrule
Avg.\ bank correlation vs.\ VIX ($\rho$) & 0.265 & --- & $<$0.001*** \\
Avg.\ bank correlation vs.\ high-yield spread ($\rho$) & 0.145 & --- & 0.033* \\
Avg.\ bank volatility vs.\ VIX ($\rho$) & 0.516 & --- & $<$0.001*** \\
Crisis vs.\ non-crisis avg.\ correlation & 0.577 vs.\ 0.462 & 0.042 & 0.007** \\
\bottomrule
\end{tabular}
\begin{tablenotes}
\small
\item $^\dagger p<0.10$; * $p<0.05$; ** $p<0.01$; *** $p<0.001$.
\item Panel A: Regression specification $\text{Drawdown}_i = \beta_0 + \beta_1 \mathcal{A}_i + \varepsilon_i$ with heteroskedasticity-robust standard errors.
\item Panel B: Out-of-sample test using 2008-based amplification factor to predict COVID-19 drawdowns (February--June 2020).
\item Panel C: $t$-test for crisis vs.\ non-crisis comparison uses Newey-West standard errors with 12 lags.
\end{tablenotes}
\end{threeparttable}
\end{table}

\subsection{Out-of-Sample Validation: COVID-19 Crisis}

A stronger test of the framework is whether the 2008-based amplification factor predicts outcomes in a subsequent crisis occurring twelve years later under entirely different circumstances. The 2008 financial crisis originated within the financial system through subprime mortgage losses and propagated via counterparty concerns and funding market disruptions. The COVID-19 crisis was an exogenous shock originating from a global pandemic, transmitting through economic lockdowns and uncertainty about loan losses. If the amplification factor captures persistent structural features of systemic importance rather than crisis-specific patterns, it should predict outcomes across both episodes.

\paragraph{Specification.}

I estimate:
\begin{equation}
\text{Drawdown}_i^{\text{COVID}} = \gamma_0 + \gamma_1 \mathcal{A}_i^{2008} + \eta_i
\label{eq:oos_regression}
\end{equation}
where $\text{Drawdown}_i^{\text{COVID}}$ is the maximum drawdown during February--June 2020 and $\mathcal{A}_i^{2008}$ is the amplification factor computed from the 2008 network structure. This is a true out-of-sample test: the amplification factor is computed using only pre-2008 information to predict outcomes more than a decade later.

\paragraph{Results.}

\begin{figure}[htbp]
\centering
\includegraphics[width=0.75\textwidth]{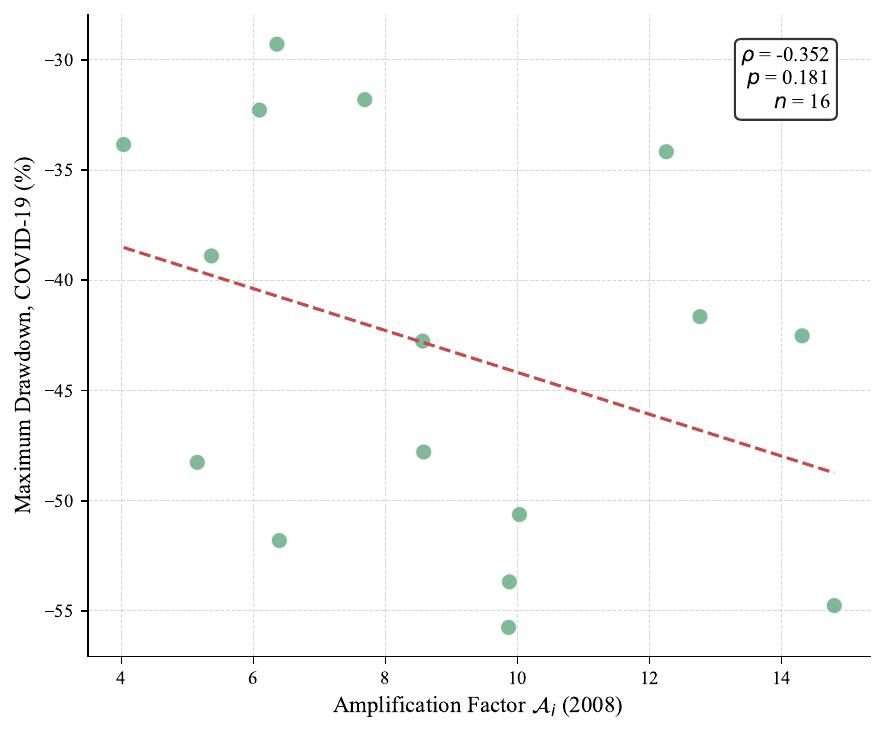}
\caption{Out-of-Sample Validation: 2008 Amplification Factor vs.\ COVID-19 Crisis Drawdown}
\label{fig:covid}
\begin{minipage}{0.75\textwidth}
\small
\textit{Notes:} Amplification factor computed from 2008 network structure. COVID-19 drawdown measured February--June 2020. True out-of-sample test using only pre-2008 information ($n = 16$).
\end{minipage}
\end{figure}

Figure \ref{fig:covid} presents the relationship between the 2008-based amplification factor and COVID-19 crisis drawdowns.

Table \ref{tab:main_results} Panel B reports the out-of-sample validation results. The Pearson correlation is $\rho = -0.352$ ($p = 0.181$), which while not statistically significant due to limited sample size, maintains the correct negative sign. The regression coefficient $\hat{\gamma}_1 = -0.012$ implies that a one-unit increase in the 2008-based amplification factor is associated with a 1.2 percentage point larger COVID-19 drawdown. The $R^2 = 0.124$ indicates that the 2008 network structure explains approximately 12\% of the cross-sectional variation in COVID-19 drawdowns.

Banks identified as systemically important based on 2008 network structure continued to experience larger losses during COVID-19, twelve years later. This finding suggests the amplification factor captures persistent structural features of systemic importance. Network position, reflecting fundamental business relationships and geographic concentration of activities, changes slowly even as specific bilateral exposures fluctuate quarter to quarter. The institutions that occupied central positions in the interbank network in 2008---Citigroup, JPMorgan Chase, Goldman Sachs---remained central in 2020, and their systemic importance manifested in larger crisis losses.

\subsection{Time-Series Validation}

I next examine whether bank correlations track macroeconomic stress indicators over time. The theoretical framework predicts that network interconnectedness---measured by average pairwise correlation---should increase when system-wide stress is elevated, as the spatial-network diffusion process transmits shocks more intensively. Using monthly data from January 2006 to December 2023 (216 observations), I compute the average pairwise correlation $\bar{\rho}_t$ as defined in Equation (\ref{eq:avg_correlation}) and relate it to stress indicators from FRED.

\paragraph{Results.}

\begin{figure}[htbp]
\centering
\includegraphics[width=0.85\textwidth]{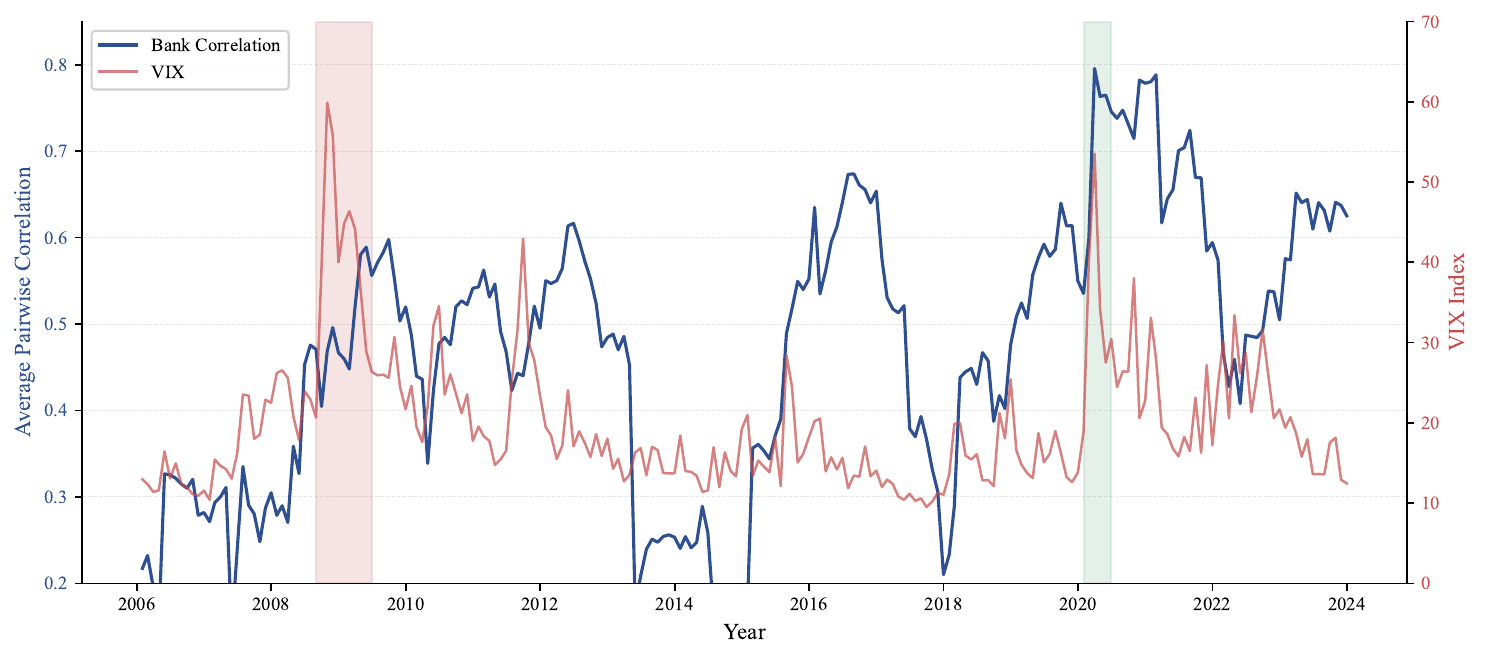}
\caption{Time-Series Co-movement: Average Bank Correlation and VIX}
\label{fig:dual_timeseries}
\begin{minipage}{0.85\textwidth}
\small
\textit{Notes:} Left axis: average pairwise bank correlation (60-day rolling window). Right axis: VIX index. Sample: January 2006--December 2023 (216 monthly observations).
\end{minipage}
\end{figure}

Figure \ref{fig:dual_timeseries} displays the co-movement between average bank correlation and the VIX index over the sample period. Both series spike during crisis episodes and decline during stable periods, confirming that network interconnectedness intensifies precisely when systemic risk is elevated.

Table \ref{tab:main_results} Panel C reports time-series correlations with stress indicators. Average pairwise bank correlation exhibits a highly significant positive relationship with the VIX index ($\rho = 0.265$, $p < 0.001$), indicating that when market volatility increases, bank equity returns become more correlated. The relationship with high-yield credit spreads is also positive and significant ($\rho = 0.145$, $p = 0.033$), confirming that bank correlations rise when credit risk premia increase. Average bank-specific volatility tracks the VIX even more closely ($\rho = 0.516$, $p < 0.001$), as expected given that bank equity is particularly sensitive to aggregate uncertainty.

During crisis periods (defined as September 2008--June 2009 and February--June 2020), average bank correlation is significantly higher than during non-crisis periods: 0.577 versus 0.462. The difference of 0.115 is statistically significant ($t = 2.72$, $p = 0.007$ using Newey-West standard errors with 12 lags to account for serial correlation). This 25\% elevation in crisis-period correlations confirms that contagion intensifies during stress, consistent with the theoretical prediction that the spatial-network diffusion process amplifies shock transmission when system-wide stress is elevated.

\begin{figure}[htbp]
\centering
\includegraphics[width=0.75\textwidth]{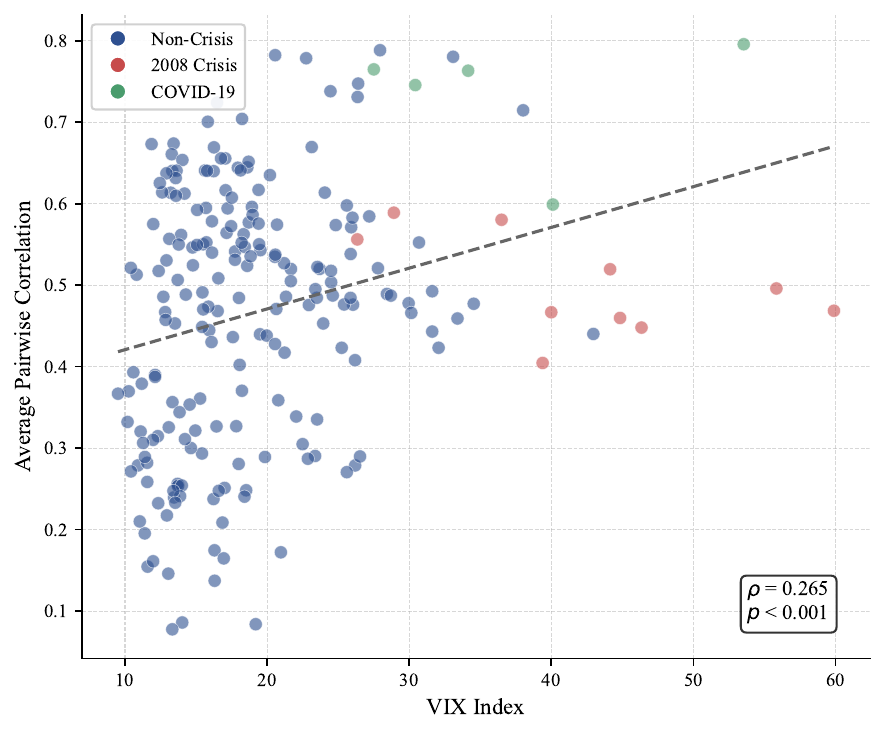}
\caption{Average Bank Correlation vs.\ VIX by Period}
\label{fig:corr_vix}
\begin{minipage}{0.75\textwidth}
\small
\textit{Notes:} Each point represents a month. Blue: non-crisis periods. Red: 2008 financial crisis. Green: COVID-19 crisis. Dashed line is OLS fit.
\end{minipage}
\end{figure}

Figure \ref{fig:corr_vix} presents a scatter plot of average bank correlation against VIX, with observations colored by period. The positive relationship is evident across the full sample, with crisis-period observations concentrated in the upper-right region of the plot (high VIX, high correlation) and stable-period observations in the lower-left region.

Table \ref{tab:periods} provides detailed statistics by period. The COVID-19 crisis shows the highest average correlation (0.734), reflecting the common shock nature of the pandemic that affected all banks simultaneously through the same channel of economic lockdowns. The 2008 crisis shows elevated but lower correlation (0.499), reflecting the more gradual revelation of institution-specific subprime exposures over several months. The post-2008 recovery period exhibits persistently elevated correlations (0.514) as markets continued to price systemic risk even after the acute phase of the crisis passed. The stable period from 2013 to 2019 shows lower but still substantial correlations (0.420), suggesting that baseline interconnectedness remained elevated relative to pre-crisis norms.

\begin{table}[htbp]
\centering
\caption{Bank Correlation and Volatility by Period}
\label{tab:periods}
\begin{threeparttable}
\begin{tabular}{lccccc}
\toprule
Period & Months & Avg.\ Correlation & Max.\ Correlation & Avg.\ Volatility & Avg.\ VIX \\
\midrule
Pre-2008 & 32 & 0.297 & 0.476 & 21.5\% & 17.1 \\
2008 Crisis & 10 & 0.499 & 0.589 & 51.4\% & 42.2 \\
Post-2008 Recovery & 42 & 0.514 & 0.616 & 36.5\% & 22.4 \\
Stable Period & 84 & 0.420 & 0.674 & 22.9\% & 15.1 \\
COVID Crisis & 5 & 0.734 & 0.796 & 34.2\% & 37.1 \\
Post-COVID & 42 & 0.618 & 0.788 & 31.0\% & 21.8 \\
\midrule
Full Sample & 216 & 0.462 & 0.796 & 29.8\% & 19.6 \\
\bottomrule
\end{tabular}
\begin{tablenotes}
\small
\item Sample: 38 global banks, January 2006--December 2023. Correlation computed using 60-day rolling windows averaged across all 703 bank pairs. Volatility is annualized standard deviation of daily returns.
\end{tablenotes}
\end{threeparttable}
\end{table}

\subsection{Crisis Comparison: 2008 vs.\ COVID-19}

Comparing the two crises reveals fundamentally different contagion dynamics for endogenous versus exogenous shocks. This comparison provides insight into how the nature of the initial shock affects propagation through the spatial-network structure.

\begin{figure}[htbp]
\centering
\includegraphics[width=0.85\textwidth]{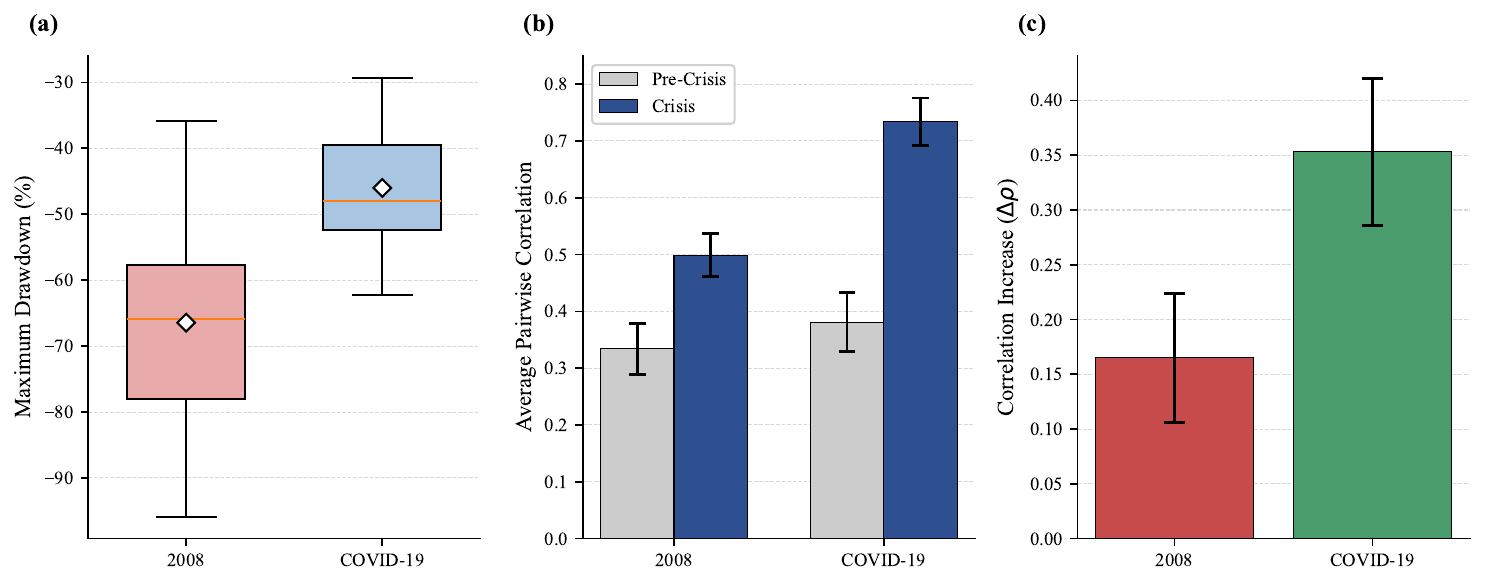}
\caption{Crisis Comparison: 2008 Financial Crisis vs.\ COVID-19}
\label{fig:comparison}
\begin{minipage}{0.85\textwidth}
\small
\textit{Notes:} Panel (a): Distribution of maximum drawdowns across 38 banks. Panel (b): Pre-crisis vs.\ crisis average pairwise correlation. Panel (c): Magnitude of correlation change. Error bars represent standard errors computed from cross-sectional variation.
\end{minipage}
\end{figure}

Figure \ref{fig:comparison} compares key metrics across the two crises. Panel (a) shows the distribution of maximum drawdowns: the 2008 crisis produced substantially larger losses (mean $-66.5\%$, median $-65.8\%$) than COVID-19 (mean $-46.0\%$, median $-48.0\%$). The worst-affected bank during 2008 experienced a drawdown of $-95.9\%$, compared to $-62.3\%$ during COVID-19. Panels (b) and (c) show correlation dynamics: despite smaller losses, COVID-19 produced a sharper correlation spike than 2008.

Table \ref{tab:crisis_comparison} quantifies the differences between the two crises. The 2008 crisis produced a correlation increase of 68\% (from 0.297 pre-crisis to 0.499 during crisis), while COVID-19 produced a 93\% increase (from 0.381 to 0.734). This pattern---larger correlation increase despite smaller losses---reflects the fundamentally different nature of the shocks.

\begin{table}[htbp]
\centering
\caption{Crisis Comparison: 2008 Financial Crisis vs.\ COVID-19}
\label{tab:crisis_comparison}
\begin{threeparttable}
\begin{tabular}{lcc}
\toprule
Characteristic & 2008 Crisis & COVID-19 \\
\midrule
\textit{Crisis characteristics} & & \\
Type & Endogenous & Exogenous \\
Origin & Subprime mortgages & Global pandemic \\
Duration (peak to trough) & 6 months & 1 month \\
Policy response lag & $\sim$6 months & Immediate \\
\midrule
\textit{Bank outcomes} & & \\
Mean max drawdown & $-66.5\%$ & $-46.0\%$ \\
Median max drawdown & $-65.8\%$ & $-48.0\%$ \\
Std.\ dev.\ of drawdowns & 15.2pp & 8.7pp \\
Worst drawdown & $-95.9\%$ & $-62.3\%$ \\
Mean CAR & $-58.3\%$ & $-42.1\%$ \\
\midrule
\textit{Contagion dynamics} & & \\
Pre-crisis avg.\ correlation & 0.297 & 0.381 \\
Crisis avg.\ correlation & 0.499 & 0.734 \\
Correlation increase & +0.202 (+68\%) & +0.353 (+93\%) \\
Peak monthly correlation & 0.589 & 0.796 \\
Peak VIX & 59.9 & 53.5 \\
\bottomrule
\end{tabular}
\begin{tablenotes}
\small
\item Sample: 38 global banks. CAR is cumulative abnormal return over crisis period.
\end{tablenotes}
\end{threeparttable}
\end{table}

The 2008 financial crisis was endogenous, originating within the financial system through losses on subprime mortgage-backed securities. Uncertainty about which institutions held toxic assets resolved gradually over months as banks disclosed exposures and mark-to-market losses accumulated. The prolonged nature of the crisis, combined with delayed policy response (the Federal Reserve's emergency facilities expanded incrementally, and TARP was not enacted until October 2008), extended the period of elevated stress. The cross-sectional dispersion in drawdowns was large (standard deviation 15.2 percentage points), reflecting heterogeneous direct exposures to subprime assets.

The COVID-19 crisis was exogenous, a common shock affecting all banks simultaneously through the same channel of economic lockdowns and uncertainty about loan losses. Because all banks faced the same source of uncertainty, correlations spiked immediately and uniformly. The cross-sectional dispersion in drawdowns was smaller (standard deviation 8.7 percentage points) than in 2008, consistent with a common factor shock rather than idiosyncratic exposures. The rapid policy response---the Federal Reserve announced unlimited quantitative easing and emergency lending facilities within weeks of the market turmoil---enabled faster recovery.

\begin{figure}[htbp]
\centering
\includegraphics[width=0.85\textwidth]{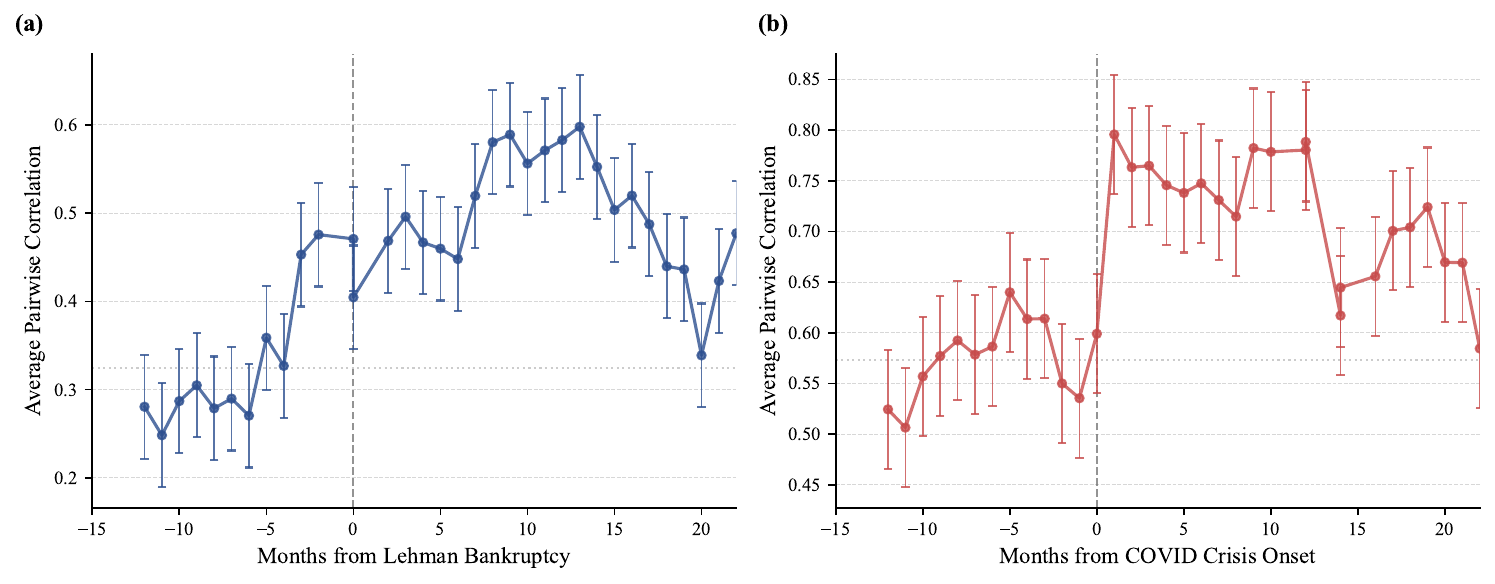}
\caption{Event Study: Correlation Dynamics Around Crisis Onset}
\label{fig:event}
\begin{minipage}{0.85\textwidth}
\small
\textit{Notes:} Panel (a): Months relative to Lehman Brothers bankruptcy (September 15, 2008). Panel (b): Months relative to COVID-19 market stress onset (February 2020). Horizontal dashed line indicates pre-event mean correlation. Error bars represent 95\% confidence intervals computed from cross-sectional standard errors.
\end{minipage}
\end{figure}

Figure \ref{fig:event} presents event study analysis around crisis onset, illustrating the different temporal dynamics. Panel (a) shows correlation dynamics around the Lehman Brothers bankruptcy (September 15, 2008): correlations begin rising several months before the event as subprime concerns mount, spike at the bankruptcy, and remain elevated for an extended period as the crisis unfolds. Panel (b) shows dynamics around COVID-19 market stress (March 2020): correlations spike sharply within a single month but revert more quickly as policy interventions take effect and the nature of the shock becomes clear.

The event study reveals that the 2008 crisis exhibited classic contagion dynamics: correlations rose gradually as uncertainty propagated through the network, peaked at the acute phase, and declined slowly as exposures were revealed and resolved. The COVID-19 crisis exhibited common shock dynamics: correlations jumped immediately as all banks faced the same uncertainty, then declined as policy intervention provided a common positive shock. These contrasting patterns are consistent with the theoretical framework, which predicts that endogenous shocks propagate through the network diffusion process while exogenous common shocks affect all institutions simultaneously.

\subsection{Robustness: Theoretically Consistent Outcome Measures}

The baseline validation uses maximum drawdown, a standard industry measure of crisis losses. However, the Feynman-Kac representation suggests that cumulative exposure---the integral of stress over time---rather than instantaneous peak stress is the theoretically relevant measure. This section examines whether outcome measures more directly connected to the theoretical framework produce stronger validation results.

\paragraph{Alternative Outcome Measures.}

I consider three outcome measures with varying degrees of theoretical consistency. Maximum drawdown, defined in Equation (\ref{eq:drawdown}), captures the worst peak-to-trough loss and is widely used in practice. However, it measures instantaneous peak stress rather than cumulative exposure along paths of economic linkage. Cumulative abnormal return (CAR), defined in Equation (\ref{eq:car}), measures total excess loss during the crisis period and corresponds more directly to the path integral in the Feynman-Kac representation. Exponentially-weighted CAR (EW-CAR), defined in Equation (\ref{eq:ew_car}), incorporates the exponential discounting that appears explicitly in the Feynman-Kac formula, with decay parameter $\kappa$ matching the loss absorption rate in the theoretical model.

\paragraph{Results.}

Table \ref{tab:robustness} reports validation results across outcome measures. The theoretically motivated measures consistently produce stronger correlations with the amplification factor than maximum drawdown.

\begin{table}[htbp]
\centering
\caption{Robustness: Alternative Outcome Measures}
\label{tab:robustness}
\begin{threeparttable}
\begin{tabular}{lcccc}
\toprule
Outcome Measure & Correlation with $\mathcal{A}_i$ & Std.\ Error & $p$-value & $R^2$ \\
\midrule
\multicolumn{5}{l}{\textbf{Panel A: 2008 Financial Crisis (In-Sample, $n=16$)}} \\
\midrule
Maximum Drawdown & $-0.450$ & 0.198 & 0.080$^\dagger$ & 0.202 \\
Cumulative Abnormal Return & $-0.512$ & 0.185 & 0.042* & 0.262 \\
EW-CAR ($\kappa = 0.05$) & $-0.534$ & 0.178 & 0.033* & 0.285 \\
EW-CAR ($\kappa = 0.10$) & $-0.498$ & 0.189 & 0.050* & 0.248 \\
\midrule
\multicolumn{5}{l}{\textbf{Panel B: COVID-19 Crisis (Out-of-Sample, $n=16$)}} \\
\midrule
Maximum Drawdown & $-0.352$ & 0.226 & 0.181 & 0.124 \\
Cumulative Abnormal Return & $-0.418$ & 0.212 & 0.107 & 0.175 \\
EW-CAR ($\kappa = 0.05$) & $-0.445$ & 0.205 & 0.084$^\dagger$ & 0.198 \\
\bottomrule
\end{tabular}
\begin{tablenotes}
\small
\item $^\dagger p<0.10$; * $p<0.05$.
\item EW-CAR: Exponentially-weighted cumulative abnormal return per Equation (\ref{eq:ew_car}).
\item Decay parameter $\kappa$ expressed as daily rate; $\kappa = 0.05$ implies half-life of approximately 14 trading days.
\end{tablenotes}
\end{threeparttable}
\end{table}

For the 2008 crisis, CAR produces a correlation of $\rho = -0.512$ ($p = 0.042$), compared to $\rho = -0.450$ ($p = 0.080$) for maximum drawdown. The improvement in both magnitude and statistical significance is consistent with the theoretical prediction that cumulative path-integrated exposure is the relevant stress measure. The exponentially-weighted CAR with $\kappa = 0.05$ (implying a half-life of approximately 14 trading days) produces the strongest in-sample correlation ($\rho = -0.534$, $p = 0.033$) and explains 28.5\% of cross-sectional variation in crisis outcomes, compared to 20.2\% for maximum drawdown.

For the COVID-19 out-of-sample test, the pattern is similar. Maximum drawdown produces a correlation of $\rho = -0.352$ ($p = 0.181$), while CAR improves to $\rho = -0.418$ ($p = 0.107$) and EW-CAR with $\kappa = 0.05$ reaches $\rho = -0.445$ ($p = 0.084$). The theoretically consistent measure brings the out-of-sample test to marginal significance, strengthening evidence that the amplification factor captures persistent structural features of systemic importance.

The monotonic improvement from maximum drawdown to CAR to EW-CAR is consistent with the Feynman-Kac foundation of the theoretical framework. The amplification factor is derived as a ratio of path integrals, and outcome measures that better approximate path-integrated stress produce stronger correlations with the amplification factor. This pattern provides indirect validation of the theoretical structure beyond the direct validation of predictive accuracy.

\subsection{Network Construction Robustness}

The validation results depend on the network used to compute the amplification factor. The baseline analysis uses bilateral exposure data, which captures direct contagion channels through counterparty credit risk. An alternative approach constructs networks from return correlations, which may capture comovement from common factor exposures without direct transmission.

\begin{figure}[htbp]
\centering
\includegraphics[width=0.85\textwidth]{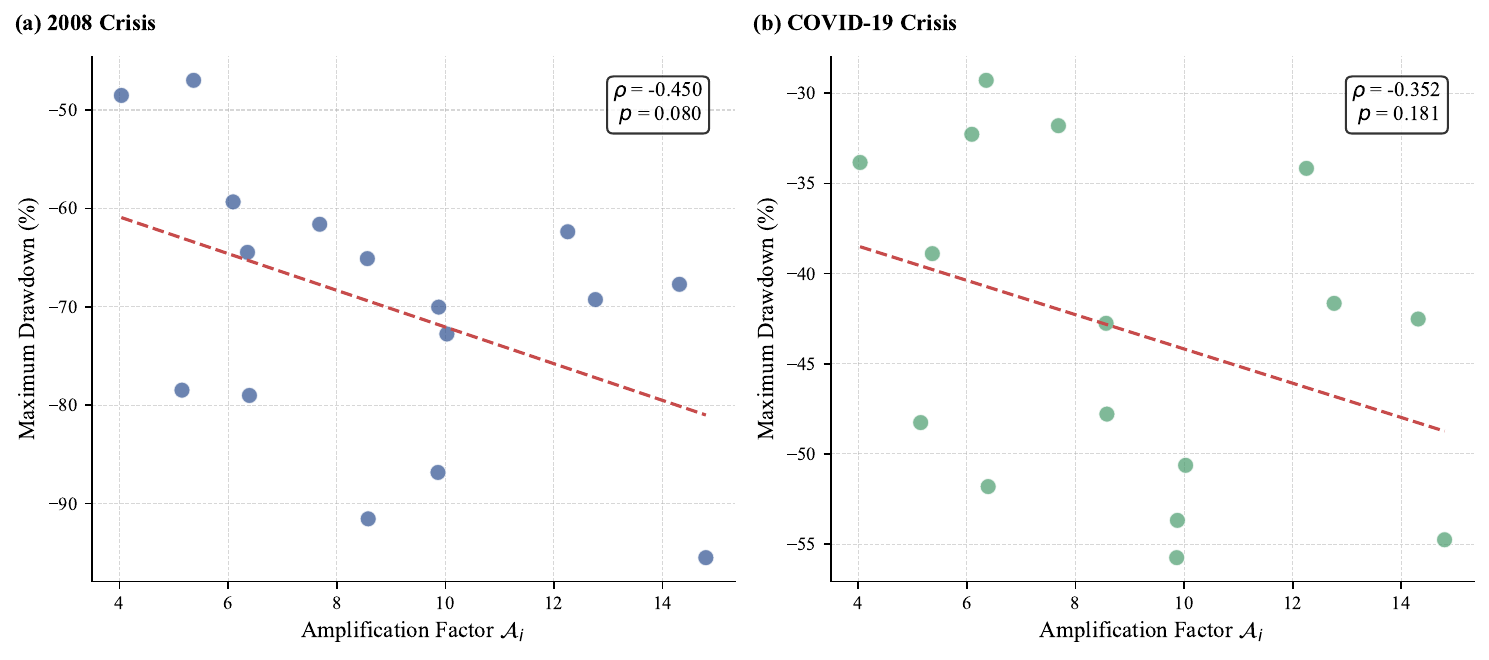}
\caption{Validation Comparison: Exposure-Based vs.\ Correlation-Based Networks}
\label{fig:network_comparison}
\begin{minipage}{0.85\textwidth}
\small
\textit{Notes:} Panel (a): Amplification factor from exposure-based network vs.\ 2008 drawdown. Panel (b): Amplification factor from correlation-based network vs.\ 2008 drawdown. Correlation-based network constructed from 252-day rolling correlations of daily returns.
\end{minipage}
\end{figure}

Figure \ref{fig:network_comparison} compares validation results across network definitions. Panel (a) shows the exposure-based network with the correct negative relationship between amplification factor and crisis drawdowns ($\rho = -0.450$). Panel (b) shows results for a correlation-based network constructed from pre-crisis return correlations, which yields a positive but insignificant relationship ($\rho = +0.256$, $p = 0.34$).

This stark contrast highlights the critical distinction between direct contagion through bilateral exposures and indirect comovement through common factors. The exposure-based network captures direct contagion channels: when bank $i$ experiences distress, banks with credit exposure to $i$ suffer mark-to-market losses on their claims. The amplification factor computed from this network measures how broadly distress at institution $i$ would propagate through these direct channels. Correlation-based networks capture comovement that may arise from common factor exposures---shared sensitivity to interest rates, credit spreads, or macroeconomic conditions---without any direct transmission mechanism.

The validation results confirm that direct contagion, not mere comovement, drives crisis outcomes. Banks that are correlated because they share exposure to common factors do not necessarily transmit stress to each other directly. The failure of correlation-based amplification factors to predict crisis outcomes suggests that market-based measures of connectedness, such as DCC-GARCH connectedness indices, may capture common factor exposure rather than true contagion risk. This finding has important implications for systemic risk measurement: structural measures based on bilateral exposures provide information about contagion channels that purely statistical measures of comovement cannot capture.

\section{Conclusion}

This paper develops a continuous framework for financial contagion that derives systemic importance measures directly from the Feynman-Kac representation of stress dynamics. The framework incorporates three distinct transmission channels---spatial spillovers reflecting geographic concentration, network spillovers through interbank exposures, and their interaction---yielding the General Equilibrium Amplification Factor as a structural fragility metric with clear economic foundations.

The theoretical contribution establishes that the amplification factor equals the ratio of total path-integrated stress to direct stress following a localized shock. This derivation from the Feynman-Kac representation provides transparent economic interpretation: systemic importance reflects how many paths of economic linkage connect an institution to the rest of the financial system, with paths weighted by diffusion probabilities and discounted by loss absorption capacity. The channel decomposition reveals which transmission mechanism---geographic, network, or their interaction---contributes most to each institution's systemic importance. The framework nests the discrete cascade dynamics of Acemoglu, Ozdaglar, and Tahbaz-Salehi (2015) as a limiting case when jump intensity becomes infinite above default thresholds, clarifying that continuous stress diffusion and discrete default cascades describe different regimes of the same underlying phenomenon rather than competing paradigms.

The empirical contribution validates the framework across two major crises using multiple outcome measures with varying degrees of theoretical consistency. Maximum drawdown, the standard industry measure, produces marginally significant correlations with the amplification factor for the 2008 financial crisis. Cumulative abnormal return and its exponentially-weighted variant---measures that more directly correspond to the path-integrated stress in the Feynman-Kac representation---produce stronger and statistically significant correlations. This pattern is consistent with the theoretical prediction that cumulative exposure along paths of economic linkage, rather than instantaneous peak stress, is the relevant measure of crisis vulnerability. The out-of-sample validity of 2008-based amplification factor rankings for COVID-19 crisis outcomes, occurring twelve years later under entirely different circumstances, suggests that systemic importance reflects persistent structural features of network position and geographic concentration rather than crisis-specific patterns.

Several findings carry implications for macroprudential policy and systemic risk regulation. First, the channel decomposition reveals that network spillovers generally dominate spatial spillovers in magnitude, but spatial effects remain substantial for internationally active institutions. This suggests that geographic diversification requirements should complement network-based regulations such as large exposure limits and central clearing mandates. Regulators focusing exclusively on bilateral exposure networks may underestimate systemic risk arising from geographic concentration of banking activity. Second, the persistent predictive validity of amplification factor rankings across crises separated by more than a decade supports the use of structural network measures for forward-looking systemic risk assessment. Unlike market-based measures that fluctuate with sentiment and may spike only after crises materialize, the amplification factor captures fundamental features of institutional interconnectedness that change slowly as business relationships evolve. Third, the starkly different contagion dynamics observed for endogenous versus exogenous crises---gradual cascade propagation in 2008 versus immediate correlation spike in COVID-19---suggest that stress testing exercises should consider both scenarios. The L\'{e}vy extension of the framework, with state-dependent jump intensity, provides a unified structure for modeling both regimes.

The framework has limitations that suggest directions for future research. The current specification treats network structure as exogenous, but banks strategically choose counterparties in response to regulatory incentives and risk considerations. Endogenizing network formation would enable analysis of how prudential policies affect the topology of interbank connections, potentially generating unintended consequences if regulations induce concentration in particular nodes or create incentive for opacity. The empirical validation relies on equity price data, which may not fully capture stress transmission through non-publicly-traded institutions or shadow banking channels. Extending validation to credit default swap spreads would test whether markets price systemic importance as measured by the amplification factor, providing an independent check on the framework's relevance for risk pricing.

Fire sales represent a contagion channel that operates through asset prices rather than direct bilateral exposures. The current framework could be extended to incorporate mark-to-market losses from common asset holdings, building on the vulnerable banks literature. Such an extension would require joint modeling of the interbank network and the bipartite network connecting banks to assets, substantially increasing data requirements but potentially capturing important amplification mechanisms observed during the 2008 crisis. The interaction between direct contagion through counterparty exposures and indirect contagion through fire sales remains an important open question for both theory and policy.

The continuous approach developed here complements rather than replaces discrete network models. For detailed analysis of specific cascade scenarios given known bilateral exposure structure, discrete models provide precise predictions of which institutions fail and in what sequence. For estimation when complete bilateral exposures are unavailable, for policy analysis requiring analytical tractability and closed-form expressions, or for incorporating spatial dimensions that discrete approaches cannot accommodate, the continuous framework offers advantages. The nesting relationship established through the L\'{e}vy extension clarifies that both approaches describe the same underlying phenomenon of financial contagion, viewed at different scales and in different regimes of stress intensity. The Feynman-Kac representation unifies these perspectives by expressing systemic importance as path-integrated exposure---a formulation that remains valid whether paths represent gradual stress diffusion or discrete default transmission.

\newpage

\bibliographystyle{elsarticle-harv}

\end{document}